# Study of Centrality Dependence of Kinetic Freeze-out Conditions in Pb + Pb Collisions at $\sqrt{s_{NN}}$= 2.76 TeV

**Saeed Uddin\*, Inam-ul Bashir and Riyaz Ahmed Bhat**

*Department of Physics, Jamia Millia Islamia (Central University)*
*New Delhi-110025*

## Abstract

The transverse momentum spectra of identified particles at midrapidity in Pb + Pb collisions at $\sqrt{s_{NN}}$ = 2.76 TeV have been studied as a function of collision centrality by using a unified statistical thermal freeze-out model. The calculated results are found to be in good agreement with the experimental data measured by the ALICE experiment at LHC. The model calculations provide the thermal freeze-out conditions in terms of the temperature and collective flow parameters for different particle species. We observe a rise in the thermal freeze-out temperature but a mild decrease in the collective flow velocity parameter from central to peripheral collisions. The model used incorporates the simultaneous effect of the longitudinal as well as transverse hydrodynamic flows. The baryon chemical potential is assumed to be zero ($\mu_B \sim 0$), a situation expected in the heavy ion collisions at LHC energies due to a high degree of nuclear transparency.

## Introduction

High-energy heavy-ion collisions at RHIC and LHC offer the unique possibility of studying the nuclear matter under the extreme conditions of temperature and pressure, in particular a possible transition to a deconfined phase of quarks and gluons called Quark–Gluon Plasma (QGP) which has been predicted by lattice QCD [1]. The transverse momentum ($p_T$) distributions and yields of identified particles are instrumental to the study of the thermal and collective flow properties of the dense and hot hadronic matter formed in such collisions. Results from lower energies [2-8] have also shown that the spectra of various hadrons emitted from the bulk matter created in high-energy nuclear reactions at the freeze-out can be quantitatively described in terms of hydrodynamic models. The initial hot and dense partonic matter (consisting of colored quarks and gluons), which may be in a plasma state, cools down due to secondary particle production and collective flow expansion during which a certain fraction of the system's thermal energy is converted into the directed flow energy due to rescattering of partons in the expanding QGP. The system undergoes a transition to a hadron gas phase through a process leading to the freezing of the color degrees of freedom [9]. This process where all quarks and gluons bind into colorless combinations of quarks and gluons is called hadronization. Gluons in the process are absorbed or fuse into various combinations of the quarks (antiquarks) thereby balancing the color charge and forming colorless hadrons [10]. The observed particle abundances in such experiments are well described in terms of thermal models. The particle momentum distributions reflect the hadronic conditions later in the evolution at the so called kinetic or thermal freeze-out of the hadron gas phase. The particle's energy and momentum spectra are frozen in time when elastic interactions cease [11]. The $p_T$ distributions carry the encoded information about the collective transverse expansion (radial flow) and the freeze-out temperature, *T*, at the kinetic freeze-out [12, 13]. The collective expansion is driven by internal pressure gradients and has been addressed within the hydrodynamic model approaches [14, 15]. The produced hadrons are believed to carry information about the collision dynamics and the subsequent space-time evolution of the system. Hence an accurate measurement of the transverse momentum distributions of identified hadrons along with the rapidity spectra is essential for the understanding of the dynamics and the properties of the created matter up to the final thermal or hydrodynamical freeze-out in case of collective flow [16].

Several successful attempts have been made to describe the transverse momentum distribution of the hadrons produced in ultra-relativistic collisions. These models incorporate the collective hydrodynamic flow along the radial (transverse) direction only. Similarly, rapidity distributions have been described successfully by incorporating a longitudinal boost (only). Recently we have proposed a model which simultaneously incorporates the longitudinal as well as transverse collective flow effects with a varying chemical potential along the rapidity axis [16], which essentially arises due to the nuclear transparency effect at the highest RHIC and the present LHC energies. It has been shown earlier [17] that this model can simultaneously explain the rapidity and transverse momentum distributions of hadrons and also their ratios in Au-Au collisions at highest RHIC energy of $\sqrt{s_{NN}}$ = 200 GeV. Also we have employed this model to successfully reproduce the transverse momentum distributions of hadrons produced in the *central* Pb + Pb collisions at $\sqrt{s_{NN}}$ = 2.76 TeV [18].

In the following we briefly describe the model and use it to reproduce the transverse momentum distributions of various hadrons produced in Pb + Pb collisions at $\sqrt{s_{NN}}$ = 2.76 TeV for different centrality classes. The main motive of doing this is to study any possible variation in the collective behaviour of the system as we go from the most central collisions to the peripheral collisions.

**Model**

The details of the model used here can be found in Reference [16]. We only briefly describe the main features of the model. The invariant cross section of hadrons emitted from within an expanding fireball, which is regarded to be in the state of a local thermal equilibrium at the time of freeze-out will have the same value in all Lorentz frames [19]. We can thus write $E \frac{d^3 n}{d^3 p} = E' \frac{d^3 n}{d^3 p'}$.

The primed quantities refer to the rest frame of the *local hadronic fluid element* while the unprimed quantities correspond to the overall rest frame of the hadronic fireball. The occupation number distribution of the hadrons in the momentum space follows the distribution function $E' \frac{d^3 n}{d^3 p'} \sim \frac{E'}{e^{(\frac{E'-\mu}{T})} \pm 1}$, where (+) and (-) signs are for fermions and bosons, respectively, and $\mu$ is the chemical potential of the given hadronic specie. For the temperatures and the large masses of baryons under consideration we can use the Boltzman distribution.

There is a strong evidence of increasing baryon chemical potential, $\mu_B$ along the collision axis in the RHIC experiments [16, 20] which is a direct consequence of the nuclear transparency effect at high energies. The chemical potential is therefore assumed to vary accordingly as $\mu_B = a + by_0^2$ [16,18,20], where $y_0$ is the rapidity of the expanding *hadronic fluid element*. It is assumed that $y_0 \propto z$ or $y_0 = \xi z$. This ensures that under the transformation $z \to -z$, we will have $y_0 \to -y_0$. This is also required to preserve the symmetry of the produced secondary hadronic matter and its collective flow about $z = 0$ along the rapidity axis in the centre of mass frame of the colliding symmetric nuclei.

The *transverse* component of the velocity of the collective flow of the hadronic fireball, $\beta_T$ is assumed to vary with the radial (transverse) coordinate $r$ in accordance with the Blast Wave model as $\beta_T(r) = \beta_T^s (\frac{r}{R})^n$ [21]. The index $n$ describes the profile of $\beta_T(r)$ in the transverse direction and $\beta_T^s$ is the hadronic fluid's *surface transverse expansion velocity*. The $\beta_T^s$ is fixed by using a parameterization $\beta_T^s = \beta_T^0 \sqrt{1 - \beta_z^2}$, where $\beta_z(z)$ represents the longitudinal velocity component of the hadronic fluid element [16]. This ensures that the net velocity $\beta$ of any fluid element given by $\sqrt{\beta_T^2 + \beta_z^2} < 1$. The transverse radius of fireball decreases following a Gaussian profile i.e. $R = r_0 \exp(-\frac{z^2}{\sigma^2})$, where $\sigma$ fixes the distribution of the hadronic matter in the transverse direction [16, 20] of the hadronic fireball formed in the overall centre of mass frame of the colliding heavy ions.

In our analysis, the contributions of various heavier hadronic resonances which decay, after the thermal freeze-out of the hadronic matter has occurred, are also taken into account [17,22]. In our study we have included only the two body decay channels which dominate the decay contributions to the hadrons considered. We have considered the decay of baryonic and mesonic resonances having masses up to 2 GeV at the temperatures considered.

**Results and Discussions**

Our model calculation results (shown by solid curves in all the cases) fit the experimental data quite well. The experimental data are taken from the ALICE Collaboration for Pb + Pb collisions at $\sqrt{s_{NN}}$ = 2.76 TeV [23]. We have shown the (statistical + systematic) errors in all the cases. We have considered the (maximum) $p_T$ range up to 5 GeV in the present analysis as was also done and discussed in [18]. The transverse momentum distributions are found to be sensitive to the values of the thermal/kinetic freeze-out temperature $T$ and the transverse flow parameter $\beta_T^0$, whereas these are found to be *insensitive* to the change in the value of $\sigma$ in our model. We have fixed the value of the parameter $\sigma$ = 5.0 [18]. The value of $\sigma$ essentially determines the *transverse size* of the hadronic matter distributed along the z-axis and has a strong effect on the shape of the rapidity spectra of the particles [16]. At the given LHC energy we expect that the midrapidity regions contain equal numbers of baryons and antibaryons [24]. Furthermore, the antiparticle/particle ratios measured at $\sqrt{s_{NN}}$ = 2.76 TeV are consistent with value ~1 at all centralities [25]. Keeping in view the above facts, we have therefore set the baryon chemical potential to be zero in our analysis. Thus we are left with temperature *T*, transverse velocity component $\beta_T^0$ and the index parameter *n* as the only (i.e. three) independent parameters. The value of *ξ* = 1 has been used for all the hadrons studied in this paper and in our previous studies as well [16, 18]. The theoretical fits for the transverse momentum spectra of all the hadrons have been normalized at the first data point (i.e. at the lowest $p_T$) to facilitate a proper comparison with the experimental data set.

In figure 1 we have shown the transverse momentum spectra of protons and antiprotons obtained from most central (0-5)% up to the most peripheral (80-90)% collisions. A good overall agreement between the theoretical model results and the experimental data is obtained for all centralities except the peripheral collision cases where slight variation is observed at high $p_T$. It is seen that for the peripheral collisions the model results do not obey the experimental data beyond 3 GeV. This is not surprising since high $p_T$ particles require more re-scatterings to thermalize but in this case they escape from the fireball before doing so. This in particular is true for more peripheral collisions where the reaction zone has a smaller size. In other words this seems to be the result of directly produced particles who either suffer no or very small number of secondary collisions and therefore escape from the fireball at a very early stage. Consequently, the data points corresponding to such particles, which are not of the thermal origin, are not seen to fall on the theoretical curve.

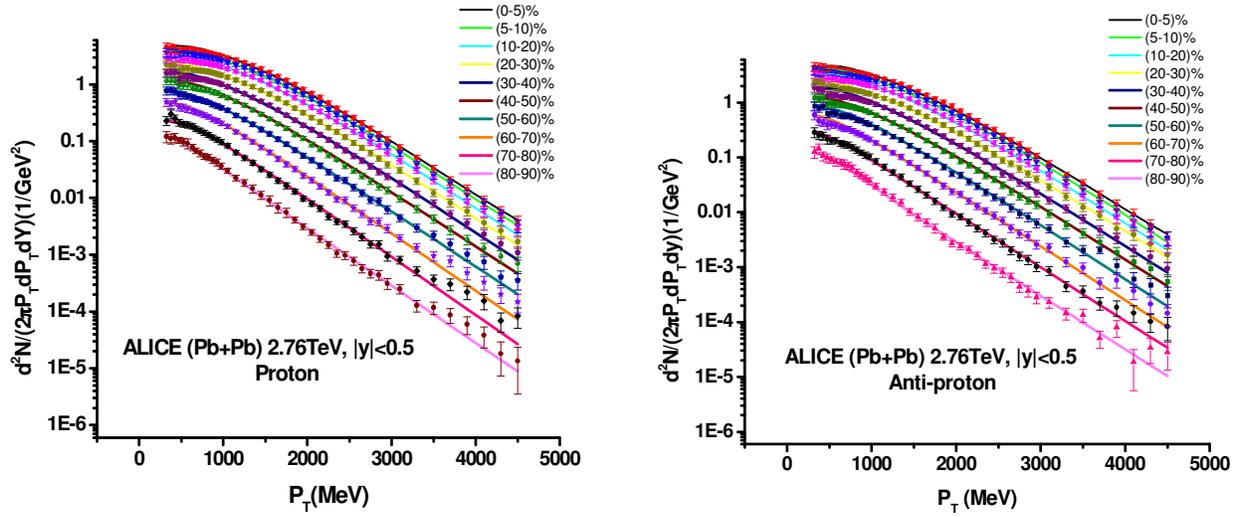

**Figure 1:** Transverse momentum spectra of protons p (left panel) and anti-protons $\bar{p}$ (right panel) for centrality classes varying from (0-5)% to (80-90)%.

The freeze-out parameters for protons for different centralities with corresponding $x^2/DoF$ are shown in Table 1. It is seen that for the protons, the transverse flow velocity parameter $\beta_T^0$ decreases from 0.88 to 0.74 whereas the thermal freeze-out temperature $T$ simultaneously increases significantly from 102.0 MeV to 172.0 MeV when going from most central (0-5)% to most peripheral (80-90)% collisions.

| Centrality (%) | (0-5) | (5-10) | (10-20) | (20-30) | (30-40) | (40-50) | (50-60) | (60-70) | (70-80) | (80-90) |
|---|---|---|---|---|---|---|---|---|---|---|
| $T$ (MeV) | 102 | 103 | 109 | 119 | 125 | 140 | 160 | 170 | 171 | 172 |
| $\beta_T^0$ | 0.88 | 0.88 | 0.87 | 0.86 | 0.85 | 0.82 | 0.78 | 0.75 | 0.74 | 0.74 |
| $n$ | 1.40 | 1.38 | 1.38 | 1.60 | 1.68 | 1.92 | 2.33 | 2.80 | 3.26 | 3.50 |
| $x^2/DoF$ | 0.80 | 0.43 | 0.45 | 0.52 | 0.34 | 0.55 | 0.32 | 0.62 | 0.68 | 0.34 |

**Table 1:** Freeze-out parameters for protons for different centrality classes with corresponding values of minimum $x^2/DoF$.

For the antiprotons (Table 2), the transverse flow velocity parameter $\beta_T^0$ decreases from 0.88 to 0.75 whereas the thermal freeze-out temperature T increases from 102.0 MeV to 171.0 MeV when going from most central (0-5)% to most peripheral (80-90)% collisions. The somewhat similar freeze-out conditions for protons and antiprotons indicate the occurrence of their near simultaneous freeze-out in the system at almost all the centralities studied.

| Centrality % | (0-5) | (5-10) | (10-20) | (20-30) | (30-40) | (40-50) | (50-60) | (60-70) | (70-80) | (80-90) |
|---|---|---|---|---|---|---|---|---|---|---|
| $T$ (MeV) | 102 | 103 | 109 | 119 | 127 | 137 | 150 | 169 | 170 | 171 |
| $\beta_T^0$ | 0.88 | 0.88 | 0.87 | 0.86 | 0.84 | 0.83 | 0.79 | 0.76 | 0.76 | 0.75 |
| $n$ | 1.40 | 1.44 | 1.56 | 1.70 | 1.97 | 2.09 | 2.58 | 3.46 | 3.47 | 3.50 |
| $x^2/DoF$ | 0.55 | 0.58 | 0.86 | 0.76 | 0.99 | 0.40 | 0.55 | 0.70 | 0.35 | 0.53 |

**Table 2: Freeze-out conditions for antiprotons for different centrality classes with corresponding values of minimum $x^2/DoF$.**

Figure 2 shows the transverse momentum spectra of K⁺ and K⁻ for different centrality classes. The experimental data are fitted quite well. Again a slight disagreement for the data points which correspond to high $p_T$ for the most peripheral cases is observed beyond 2.5 GeV which is expected in the framework of a thermal model as discussed above.

The freeze-out conditions for the different centrality classes in case of Kaons K⁺ and antiKaons K⁻ are shown in Table 3 and Table 4 respectively.

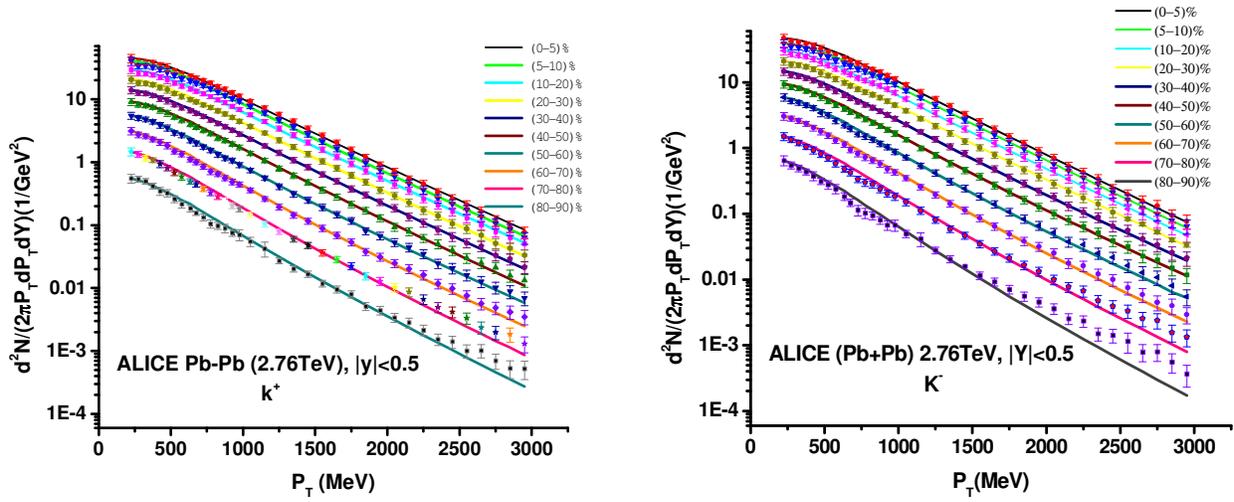

**Figure 2: Transverse momentum distributions of Kaons $K^+$ (left panel) and antiKaons $K^-$ (right panel) for centrality classes (0-5)% to (80-90)%.**

Again an increase in thermal freeze-out temperature and a decrease in the transverse flow velocity parameter is observed when going from most central (0-5)% to most peripheral (80-90)% collisions.

| Centrality % | (0-5) | (5-10) | (10-20) | (20-30) | (30-40) | (40-50) | (50-60) | (60-70) | (70-80) | (80-90) |
|---|---|---|---|---|---|---|---|---|---|---|
| $T$ (MeV) | 103 | 108 | 110 | 113 | 122 | 130 | 145 | 160 | 171 | 173 |
| $\beta_T^0$ | 0.89 | 0.88 | 0.88 | 0.87 | 0.86 | 0.84 | 0.82 | 0.79 | 0.76 | 0.74 |
| $n$ | 1.80 | 1.98 | 1.98 | 1.99 | 2.17 | 2.38 | 2.55 | 2.81 | 2.95 | 3.0 |
| $x^2/DoF$ | 0.34 | 0.54 | 0.31 | 0.30 | 0.21 | 0.29 | 0.21 | 0.68 | 1.20 | 1.14 |

**Table 3: Freeze-out conditions for Kaons $K^+$ for different centrality classes with corresponding values of minimum $x^2/DoF$.**

| Centrality % | (0-5) | (5-10) | (10-20) | (20-30) | (30-40) | (40-50) | (50-60) | (60-70) | (70-80) | (80-90) |
|---|---|---|---|---|---|---|---|---|---|---|
| $T$ (MeV) | 105 | 105 | 108 | 112 | 132 | 142 | 146 | 152 | 161 | 172 |
| $\beta_T^0$ | 0.88 | 0.88 | 0.88 | 0.87 | 0.84 | 0.82 | 0.79 | 0.77 | 0.75 | 0.73 |
| $n$ | 1.80 | 1.80 | 1.97 | 1.98 | 2.51 | 2.66 | 2.70 | 2.94 | 2.98 | 3.0 |
| $x^2/DoF$ | 0.34 | 0.33 | 0.37 | 0.50 | 0.29 | 0.26 | 0.93 | 1.06 | 2.15 | 3.16 |

**Table 4: Freeze-out conditions for antiKaons K⁻ for different centrality classes with corresponding values of minimum $x^2/DoF$.**

The transverse momentum spectra of lambda $\Lambda$ and $K_S^0$ are shown in figure 3. The experimental data are again fitted quite well except for most peripheral cases where a deviation is observed beyond $p_T = 3.5$ GeV. Again this is clearly due to the small reaction volume formed at these centralities wherein the high $p_T$ particles particularly don't get enough time to thermalize.

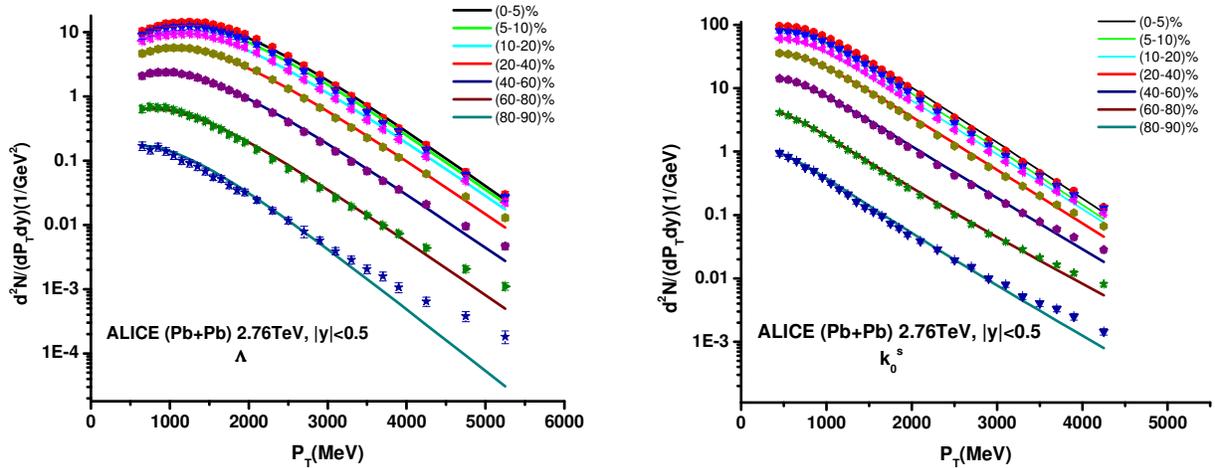

**Figure:3 Transverse momentum distribution of $\Lambda$ (left panel) and $K_S^0$ (right panel) from centrality classes (0-5)% to (80-90)%.**

The freeze-out conditions of these particles for different centrality classes are shown in table 5 and table 6, respectively.

| Centrality % | (0-5) | (5-10) | (10-20) | (20-40) | (40-60) | (60-80) | (80-90) |
|---|---|---|---|---|---|---|---|
| $T$ (MeV) | 127 | 127 | 133 | 140 | 150 | 169 | 174 |
| $\beta_T^0$ | 0.84 | 0.84 | 0.83 | 0.82 | 0.79 | 0.73 | 0.70 |
| $n$ | 1.06 | 1.06 | 1.10 | 1.20 | 1.40 | 1.82 | 1.96 |
| $x^2/DoF$ | 0.52 | 0.52 | 0.60 | 0.85 | 2.0 | 2.80 | 3.20 |

**Table:5 Freeze-out conditions for lambda $\Lambda$ for different centrality classes with corresponding minimum $x^2/DoF$.**

| Centrality % | (0-5) | (5-10) | (10-20) | (20-40) | (40-60) | (60-80) | (80-90) |
|---|---|---|---|---|---|---|---|
| $T$ (MeV) | 125 | 127 | 139 | 140 | 150 | 161 | 173 |
| $\beta_T^0$ | 0.84 | 0.84 | 0.84 | 0.82 | 0.82 | 0.77 | 0.72 |
| $n$ | 1.61 | 1.63 | 1.67 | 1.80 | 2.04 | 2.12 | 2.30 |
| $x^2/DoF$ | 1.65 | 1.70 | 1.80 | 2.10 | 2.34 | 2.50 | 3.02 |

**Table:6 Freeze-out conditions for $K_S^0$ for different centrality classes with corresponding minimum $x^2/DoF$.**

The transverse momentum spectra of cascade $\Xi^-$ and anticascade $\overline{\Xi}$ are shown in figure 4. A very good overall fit is obtained for the whole centrality classes considered up to $p_T \sim 5$ GeV.

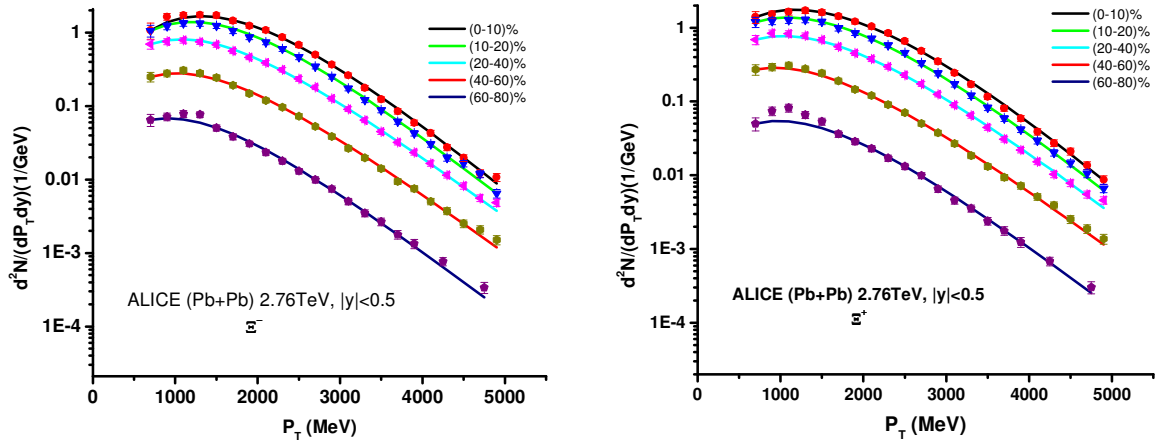

**Figure 4: Transverse momentum distribution of cascade $\Xi^-$ (left panel) and anticascade $\overline{\Xi}$ (right panel) for centrality classes varying from (0-10)% to (60-80)%.**

The different freeze-out parameters for cascade $\Xi^-$ and anticascade $\overline{\Xi}$ are summarized in table 7 and table 8, respectively.

| .Centrality % | (0-10) | (10-20) | (20-40) | (40-60) | (60-80) |
|---|---|---|---|---|---|
| $T$ (MeV) | 133 | 148 | 160 | 169 | 175 |
| $\beta_T^0$ | 0.81 | 0.79 | 0.77 | 0.75 | 0.70 |
| $n$ | 0.90 | 1.23 | 1.45 | 1.57 | 1.64 |
| $x^2/DoF$ | 0.38 | 0.57 | 0.90 | 0.98 | 1.60 |

**Table 7: Freeze-out conditions for cascade $\Xi^-$ for different centrality classes with corresponding minimum $x^2/DoF$.**

The transverse momentum spectra of omega $\Omega$ and anti-omega $\overline{\Omega}$ are shown in figure 5. Here again a good agreement is seen between the theoretical results and the experimental data points up to $p_T = 5$ GeV.

| Centrality % | (0-10) | (10-20) | (20-40) | (40-60) | (60-80) |
|---|---|---|---|---|---|
| $T$ (MeV) | 149 | 158 | 164 | 169 | 175 |
| $\beta_T^0$ | 0.80 | 0.78 | 0.76 | 0.74 | 0.69 |
| $n$ | 1.25 | 1.43 | 1.53 | 1.60 | 1.65 |
| $x^2/DoF$ | 0.48 | 0.49 | 0.90 | 0.91 | 1.79 |

**Table 8: Freeze-out conditions for anticascade $\overline{\Xi}$ for different centrality classes with corresponding minimum $x^2/DoF$.**

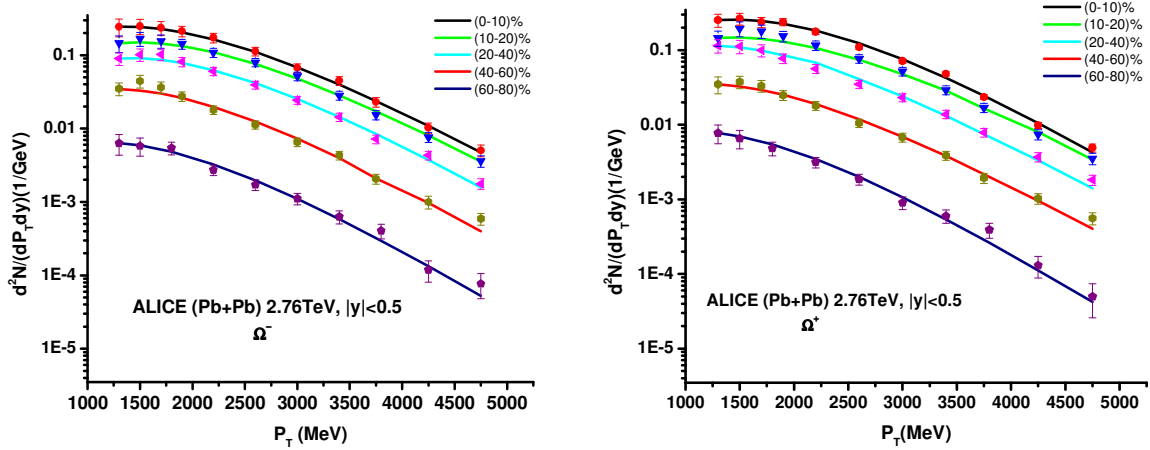

**Figure 5: Transverse momentum distribution of omega $\Omega$ (left panel) and anti-omega $\overline{\Omega}$ (right panel) from centrality classes varying from (0-10)% to (60-80)%.**

The freeze-out conditions for these particles are summarized in table 9 and table 10, respectively.

We have also compared our model results in terms of freeze-out parameters with those obtained in Au + Au collisions at highest RHIC energy of $\sqrt{s_{NN}}$ = 200 GeV [17] for the case of protons produced in the collisions in figure 6.

| Centrality % | (0-10) | (10-20) | (20-40) | (40-60) | (60-80) |
|---|---|---|---|---|---|
| $T$ (MeV) | 155 | 157 | 163 | 169 | 176 |
| $\beta_T^0$ | 0.77 | 0.78 | 0.76 | 0.73 | 0.69 |
| $n$ | 1.22 | 1.21 | 1.21 | 1.23 | 1.29 |
| $x^2/DoF$ | 0.10 | 0.52 | 0.86 | 1.0 | 1.13 |

Table 9: Freeze-out conditions for omega $\Omega$ for different centrality classes with corresponding values of minimum $x^2/DoF$.

| Centrality % | (0-10) | (10-20) | (20-40) | (40-60) | (60-80) |
|---|---|---|---|---|---|
| $T$ (MeV) | 154 | 158 | 163 | 170 | 175 |
| $\beta_T^0$ | 0.77 | 0.77 | 0.75 | 0.74 | 0.69 |
| $n$ | 1.23 | 1.22 | 1.26 | 1.27 | 1.28 |
| $x^2/DoF$ | 0.20 | 0.41 | 1.10 | 1.0 | 1.0 |

Table 10: Freeze-out conditions for anti-omega $\overline{\Omega}$ for different centrality classes with corresponding minimum $x^2/DoF$.

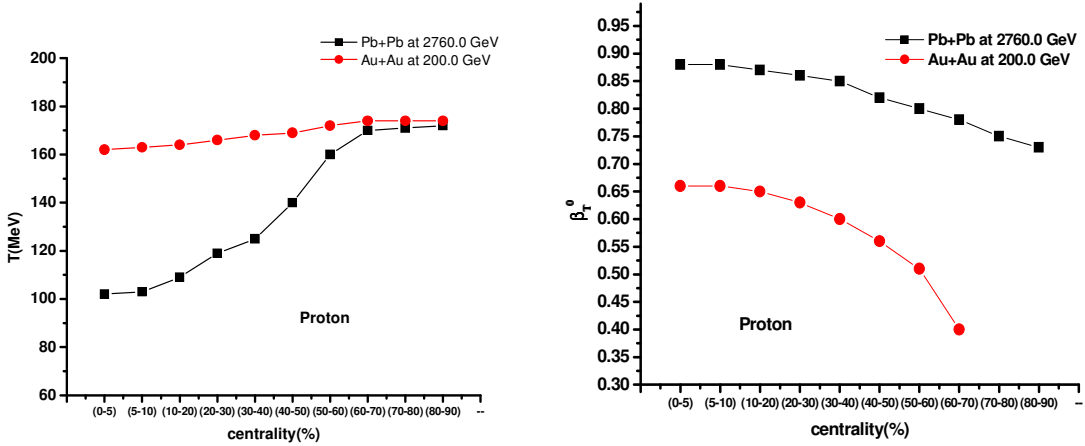

Figure: 6 Thermal freeze-out temperature T (left) and transverse flow parameter $\beta_T^0$ (right) for protons plotted as a function of collision centrality in Au + Au collisions at $\sqrt{s_{NN}}$ = 200 GeV (Red filled circles) and Pb+Pb collisions at $\sqrt{s_{NN}}$ = 2.76 TeV (Black filled squares).

We observe stronger flow present in the system at LHC than at RHIC at all centralities, which seems to be due to more particle production and hence the larger system size at LHC. Simultaneously at LHC the thermal freeze-out temperature drops rapidly below than that at RHIC particularly at higher centralities. This seems to arise due to a larger fraction of the available thermal energy being converted into directed hydrodynamic flow at LHC due to multiple collisions occurring in the dense hadronic matter spanning a larger volume at freeze-out. The same is responsible for the increase in temperature and a steady drop in $\beta_T^0$ at LHC as the centrality *decreases*.

**Summary and Conclusion**

In summary, the transverse momentum spectra of the hadrons namely p, $\bar{p}$, $K^+$, $K^-$, $K_S^0$, $\Lambda$, $\Omega$, $\bar{\Omega}$, $\Xi^-$ and $\overline{\Xi}$ produced in Pb + Pb collisions at all collision centralities are analyzed at the LHC energy $\sqrt{s_{NN}}$ = 2.76 TeV. These spectra are well described by using our earlier proposed unified thermal freeze-out model for all centrality classes. The overall model predictions agree quite well with the experimental data. However some discrepancies are seen at large impact parameters and particularly at large transverse momenta $p_T \gtrsim$ 3.0 GeV. This is understood as the high $p_T$ particles require more rescatterings to thermalize but escape from the fireball at an early stage. This effect is in particular more prominent in the peripheral collisions wherein a smaller reaction volume exists. We also observe an earlier freeze-out of hyperons as compared to lighter mass particles i.e. Kaons and protons in all centrality cases. The reason for this can be attributed to an early freeze-out for the massive particles (hyperons) when the thermal temperature is high and the collective flow is in the early stage of development and consequently $\beta_T^0$ is small. The early freeze-out of these particles is due to their *smaller cross-section with the hadronic matter*. This can also be understood in terms of the mean free path, λ, of a particle in a thermal environment which is given by λ = 1/νρ, where ν is the mean thermal cross-section of the particle with the surrounding matter having density ρ. Clear evidence is found of the dependence of thermal freeze-out temperature on the centrality of the collision with a rise in temperature from central to peripheral collisions. The transverse flow velocity parameter is simultaneously found to decrease from central to peripheral collisions. The assumption of vanishing chemical potential at midrapidity is supported by the antiproton to proton ratio which is nearly unity for all centralities. The effect of almost complete transparency in Pb + Pb collisions at LHC energy of 2.76 TeV is therefore evident.

## Acknowledgements

Inam-ul Bashir is thankful to the University Grants Commission (UGC) for awarding the Basic Scientific Research (BSR) Fellowship. Riyaz Ahmed Bhat is grateful to Council of Scientific and Industrial Research (CSIR), New Delhi for awarding Senior Research Fellowship (SRF). Saeed Uddin is grateful to the University Grants Commission (UGC) for financial assistance.